\documentstyle[preprint,eqsecnum,aps,epsf]{revtex}

\begin{document}
\tightenlines
\preprint{\vbox{
\hbox{INPP-UVA-01-03} 
\hbox{hep-ph/0106105}
}}
\draft
\def\be{\begin{eqnarray}}
\def\en{\end{eqnarray}}
\def\non{\nonumber}
\def\la{\langle}
\def\ra{\rangle}
\def\up{\uparrow}
\def\dw{\downarrow}
\def\ep{\varepsilon}
\def\ms{\overline{\rm MS}}
\def\ums{{\mu}_{_{\overline{\rm MS}}}}
\def\u{\mu_{\rm fact}}
\def\pr{{\sl Phys. Rev.}~}
\def\ijmp{{\sl Int. J. Mod. Phys.}~}
\def\epj{{\sl Eur. Phys. J.}~}
\def\jp{{\sl J. Phys.}~}
\def\mpl{{\sl Mod. Phys. Lett.}~}
\def\nc{{\sl Nuovo Cimento}~}
\def\prp{{\sl Phys. Rep.}~}
\def\prl{{\sl Phys. Rev. Lett.}~}
\def\pl{{\sl Phys. Lett.}~}
\def\np{{\sl Nucl. Phys.}~}
\def\ppnp{{\sl Prog. Part. Nucl. Phys.}~}
\def\arnps{{\sl Ann. Rev. Nucl. Part. Sci.}~}
\def\zp{{\sl Z. Phys.}~}

\title{\bf TESTS OF SU(3) SYMMETRY-BREAKING FOR\\
BARYONIC BETA TRANSITIONS IN THE STANDARD MODEL $^*$
\\}

\author{P.~K.~Kabir and X.~Song\\}

\address{Institute of Nuclear and Particle Physics\\
Jesse W. Beams Laboratory of Physics\\
Department of Physics, University of Virginia\\
Charlottesville, VA 22901, USA}

\maketitle
\vskip 3pt
\begin{abstract}
It is generally assumed that deviations from flavor SU(3) symmetry
arise entirely from quark mass-differences,
reflected in the mass splittings between strange and nonstrange
members of the same SU(3) multiplet. Under this assumption, a 
parametrization is proposed which expresses the ratios of Gamow-Teller
to Fermi matrix elements for nucleon and hyperon beta decays entirely
in terms of two SU(3)-invariant coupling constants $F$ and $D$ and two
parameters $\gamma$ and $\delta$ representing SU(3) breaking effects.
Therefore, in principle, measurement of this ratio for any four 
beta-transitions should yield all four parameters. Available data do not 
show any evidence for SU(3) breaking. Improved measurements, also for
transitions not previously measured, would provide more stringent tests.
\end{abstract}
\bigskip
\bigskip
\bigskip
\pacs{13.30.Ce,~11.30.Hv,~14.20.Jn,~23.40.-s\\
$^*$~ A preliminary version was presented at the Seattle Conference on
Symmetries in Particle Physics, June, 1997.}

\newpage

\leftline{\bf Introduction}
\smallskip
    
In the Standard Model, based on gauge-interactions of leptons and quarks, 
if one ignores the influence of more massive quarks belonging to the 
third family, charge-changing weak interactions of strange and non-strange 
hadrons are induced by the current $\bar u\gamma_{\alpha}(1+\gamma_5)d_c$ 
with

$$  d_c=d\cdot{\rm cos}\theta+s\cdot{\rm sin}\theta,          
\eqno (1)$$
where the particle symbols represent the quark fields  and the single 
parameter \cite{gl60,cab63} $\theta$ characterizes the intrinsic 
difference in strength between the weak couplings of strangeness-changing 
and strangeness-conserving currents. Had $d$ and $s$ quarks been degenerate 
in mass, $d_c$ and its orthogonal partner $s_c$ would have been equally 
valid choices for the basis states $(d,s)$ of charge-(-1/3) quarks, and 
there would have been no need to introduce $\theta$ at all. From this 
viewpoint, the charged weak current transforms like a SU(2) generator, 
as it would be required to if it solely induced $d\leftrightarrow u$
transitions. However, for reasons not yet understood, it operates in a 
SU(2) subspace [of the SU(3) space spanned by $(u,d,s)$ states] which 
is {\it skewed} with respect to the SU(2) space spanned by the (nearly) 
degenerate $(u,d)$ states \cite{note1}. $\Delta$S=1 weak interactions 
occur only because the direction in (flavor)- SU(3) space chosen by the 
charged weak current, lies outside the subspace which is orthogonal to 
the direction containing the more massive $s$ quark.
      
The description of baryonic beta-transitions in this theory requires, 
in addition to the parameter $\theta$ mentioned above  - which we shall 
take as already determined from comparison of strange and non-strange 
{\it meson} decays - an additional parameter to determine the relative 
contribution of two different ways to construct a SU(3)-octet 
matrix-element between the initial and final baryon octet states, even 
in the limit of exact SU(3) symmetry \cite{cab63}. In the Standard Model, 
this parameter $F/D$ would be determined by the quark structure of the 
baryon states which, regrettably, is only poorly known, basically 
because we do not yet have a reliable way to calculate hadronic 
strong-interaction effects. For this reason, empirical knowledge of 
this parameter is of great interest. The value deduced from global 
analyses of hyperon and nucleon beta-decays has been combined \cite{emc88} 
with measurements of deep inelastic electron or muon scattering from 
polarized nucleon targets, to make inferences about the distribution 
of spin within the nucleon. Because the deduced distributions do not 
conform to theoretical expectations, various \cite{cheng-ji,song98} 
proposals have 
been advanced to resolve this so-called ``spin crisis''. Among these 
is the suggestion \cite{es95} that the effects of SU(3) symmetry-breaking, 
arising from quark mass-differences, may be so large as to invalidate 
the values of $F/D$ deduced from the analysis of hyperon beta-decays, 
which did not take into account such effects. The aim of this brief 
report is to present an analysis of baryonic beta-transitions, 
$including$ effects of mass-differences, extending the earlier 
perturbative discussions \cite{ag64,note2}, which appears to be more 
convenient for the present purpose. Applied to the available data, we 
find no evidence for significant departure from the SU(3)-symmetric 
results, in agreement with our results \cite{skm96} from a purely 
perturbative analysis published earlier. Four relatively 
well-measured baryonic beta-transitions permit a unique determination 
of the parameters $F$ and $D$, and two additional parameters describing 
SU(3)-breaking effects, in our approach. 

A clear test of our suggested 
method would be to see if $other$ as yet poorly measured beta transitions 
of hyperons are well-described by the {\it same} set of four parameters. 
Section 2 outlines our method of analysis. The assumptions underlying 
our calculation are stated, and the resulting formula for $g_A/g_V$ 
for baryonic beta-transitions is worked out for the cases of interest. 
By applying this to the available data, we find the best values for $F$ 
and $D$; two additional parameters, representing SU(3)-breaking effects 
in our method of analysis, are found to be not distinguishable from zero, 
within the reported errors of measurement. Our conclusions are summarized 
in Section 3.
\bigskip

\leftline{\bf 2. Baryonic Beta-Transitions including SU(3) Breaking}

We shall assume that beta-transitions between octet baryons are 
induced by vector and axial-vector charged weak currents, coupled to 
corresponding V-A currents of leptons. In the Standard Model, 
these currents involve transitions between the basic quarks, and 
therefore transform as (flavor)-SU(3) octet operators. Under the 
assumption that all deviations from flavor-SU(3) symmetry arise from 
differences between the masses of strange and non-strange quarks 
\cite{note1}, the matrix-elements of the weak current will transform in 
the same way under charge-conjugation as they would in the symmetric 
limit. With the Cabibbo current:
$$j_W^{\mu}={\bar q}{{\lambda_W}\over 2}\gamma^{\mu}(1+\gamma_5)q
\eqno (2)$$          
where
$$ \lambda_W=(\lambda_1+i\lambda_2){\rm cos}\theta +
(\lambda_4+i\lambda_5){\rm sin}\theta
\eqno (3)$$          
and $\theta$ is Cabibbo's angle and the $\lambda$-matrices are defined
in \cite{gl60,pdg00}, the generalization of the assumption that 
``first-class'' 
currents \cite{weinberg58} are the only ones that occur, is
$$                  
(G'P)J_{\alpha}(G'P)^{-1}= - J_{\alpha}         
\eqno (4)$$          
where $G'=C.e^{i\pi I'_2}$ with 
$I'_2 = F_2{\rm cos}\theta+F_5{\rm sin}\theta$, 
as would be the case in the Standard Model.  
\smallskip
In the allowed approximation \cite{note3}, which suffices to describe the 
existing data, the general form of the baryonic matrix-elements, linear 
in the weak current and in $\mu$ (defined below), with the stated 
transformation properties, is
\smallskip
$$ M=a||\lambda_W\bar{\bf B}{\bf B}||+
b||\lambda_W{\bf B}\bar{\bf B}||+
c(||\lambda_W{\bf B}\mu\bar{\bf B}||-||\lambda_W\bar{\bf B}\mu{\bf B}||)+
d(||\lambda_W{\bf B}||\cdot||\mu\bar{\bf B}||+||\lambda_W\bar
{\bf B}||\cdot||\mu{\bf B}||)
\eqno (5)$$          
where 
$|| X ||\equiv$ Tr(X) and {\bf B} is the matrix representing the 
baryon octet:

$${\bf B}=\pmatrix{ 
{1\over {\sqrt 2}}\Sigma^0+{1\over {\sqrt 6}}\Lambda^0&\Sigma^+&p\cr
\Sigma^-&-{1\over {\sqrt 2}}\Sigma^0+{1\over {\sqrt
6}}\Lambda^0&n\cr
\Xi^-&\Xi^0&-{2\over {\sqrt 6}}\Lambda^0\cr}.
\eqno (6)$$          
\bigskip

In the SU(3) symmetric limit, the only possible SU(3)-octet 
matrix-elements between octet baryons are those given by the first two 
terms in Eq.(5). The effects of SU(3)-symmetry breaking are included in 
the terms containing 
$$\mu =(I-{\sqrt 3}\lambda_8)/3.
\eqno (7)$$ 
The use of the idempotent operator $\mu$ instead of $\lambda_8$ offers the 
advantage that higher powers are automatically included \cite{note4}.  
\smallskip                
  
It is known \cite{bs60,ag64} that the effects of SU(3)-breaking, 
represented 
here by the operator $\mu$, do not appear in first order for the vector 
coupling, and we shall disregard them accordingly. If we extend to the 
Gell-Mann-Le'vy-Cabibbo current, Eq.(2), the notion of the 
Conserved Vector Current, then the general form of the vector ( Fermi )
matrix-element will be further constrained by the requirement $b = -a$ in
Eq. (5).  With these assumptions, Eq.(5) yields, for the ratio $r$ of Gamow-
Teller \cite{note5} to Fermi matrix-elements for the four best-measured 
beta-
transitions \cite{pdg00}:
\smallskip
$$r_1=(-g_A/g_V)_{n\to p}=F+D+\gamma,
\eqno (8a)$$   
$$r_2=(-g_A/g_V)_{\Lambda\to p}=F+D/3+2(\gamma+\delta)/3,   
\eqno (8b)$$   
$$r_3=(-g_A/g_V)_{\Sigma^-\to n}=F-D,
\eqno (8c)$$   
$$r_4=(-g_A/g_V)_{\Xi^-\to \Lambda}=F-D/3+2(\gamma-\delta)/3.  
\eqno (8d)$$   
\smallskip
It will be seen that these formulas require, in addition to the two 
SU(3) parameters $F$ and $D$, called for in the symmetric limit, only 
two further parameters \cite{note6} $\gamma$ and $\delta$ to describe the 
effects of SU(3)-breaking. In principle, all four parameters are 
uniquely determined by the four $g_A/g_V$ ratios.  

From these equations, we obtain the following relations between $F$, $D$,
and the two additional parameters $\gamma$, and $\delta$, required to 
describe SU(3) breaking, and the reported $g_A/g_V$ ratios:
$$F=2(r_1+r_3)-{3\over 2}(r_2+r_4),
\eqno (9a)$$
$$D=2r_1+r_3-{3\over 2}(r_2+r_4),
\eqno (9b)$$
$$\gamma=3[(r_2+r_4)-(r_1+r_3)],
\eqno (9c)$$
$$\delta={1\over 2}(3r_2-2r_1-r_3).
\eqno (9d)$$
\smallskip
The Gell-Mann-Le'vy-Cabibbo hypothesis, - that the weak currents should 
transform as SU(3) octets, - would require that, in the absence of SU(3) 
breaking effects, each of the two parameters $\gamma$ and $\delta$ should 
vanish. Inserting the current values \cite{pdg00} for the $r_j$ ( $j=1-4$ 
),
$$r_1=1.2670 \pm 0.0035,   
\eqno (10a)$$
$$r_2=0.718 \pm 0.015,   
\eqno (10b)$$
$$r_3=-0.340 \pm 0.017,   
\eqno (10c)$$
$$r_4=0.25 \pm  0.05,
\eqno (10d)$$
\smallskip 
in Eqs.(9a)-(9d), we find  
$$F = 0.402 \pm 0.085,   
\eqno (11a)$$
$$D = 0.742 \pm 0.080,   
\eqno (11b)$$
$$\gamma = 0.123 \pm 0.164,   
\eqno (11c)$$
$$\delta = -0.020 \pm  0.024,
\eqno (11d)$$
\smallskip
which shows that, within the stated errors, there is no significant 
evidence for the SU(3)- breaking effects suggested by some authors[7]. 
It will be seen that the parameter $\delta$ is determined with 
relatively high precision and imposes quite a strict limit on possible 
SU(3) breaking. Because of the large error associated with the reported 
value (10d), for the ratio of $g_A/g_V$ for the $\Xi^-\to\Lambda$ 
transition, the determinations (11a)$-$(11c) for $F$, $D$, and 
$\gamma$ from Eqs. (9a)$-$(9c) are subject to correspondingly larger 
uncertainties. In particular, the value of $F/D$ deduced from Eqs. 
(11a) and (11b), while consistent with the values given by earlier
investigations \cite{close93,rat99} and used in the nucleon spin-structure 
analyses \cite{e143-smc} [assuming exact SU(3) symmetry], has an
uncertainty which is an order of magnitude larger than the error
quoted by the cited authors. Although our calculations do not fully 
support their optimistic estimation of error, uncertainties of $F$ 
and $D$ smaller than those quoted in Eqs. (11a) and (11b) could be 
justified as follows.

Eqs. (11) show that errors associated with the quantities 
$F$, $D$, and $\gamma$ are much larger than those for $\delta$. 
They arise almost entirely from the poorly determined ratio $r_4$, 
Eq. (10d), of $g_A/g_V$ for $\Xi^-\to\Lambda$ transitions, which does 
{\it not} enter the calculation, Eq. (9d), for $\delta$. Using 
available data, we have shown that the SU(3)-breaking parameter 
$\delta$ differs from zero by less than one standard deviation. The 
following remarks can be made:
\begin{itemize}
\item{If we were to accordingly set $\delta=0$, then only three 
parameters would remain to be determined.  
The first impulse, to use the more accurate data for $r_1$, $r_2$, and 
$r_3$ to solve for $F$, $D$ and $\gamma$, turns out to be unfeasible. 
Eqs. (8a)$-$(8c) are no longer linearly independent if $\delta=0$, as 
shown explicitly in Eq. (9d). Therefore, one $must$ include $r_4$ to 
solve for $F$, $D$ and $\gamma$. While it is a matter of choice 
whether one chooses $r_2$ or $r_3$ as the redundant datum, the solution
obtained by solving (8a),(8c), and (8d) yields a smaller error for $F$ 
than if (8a), (8b), and (8d) were used \cite{note7}. The first set of 
results for $F$, $D$, and $\gamma$ is shown, together with a least-squares 
fit to all four Eqs.(8a)-(8d), in the second column of Table I.}
\item{If the SU(3)-breaking parameter $\gamma$ is $assumed$ to be 
similarly small, then an analysis neglecting both  $\gamma$ and 
$\delta$ requires only 2 $r$-values to determine $F$ and $D$ 
individually. Using the relatively well-measured $g_A/g_V$ ratios 
($r_1$, $r_3$) [rather than ($r_1$, $r_2$) \cite{note7}] reported in 
Eqs. (10a) and (10c), the solution is listed on the right side of 
the third column of Table I, and compared to a least-squares fit to 
all four $r$-values.}
\item{The linear dependence of Eqs. (8a)-(8c), when $\delta$ is 
$assumed$ to be zero, suggests the following approach to their 
solution. By combining these equations, we find 
$$     r_1 - r_3 = 2 D + \gamma = 1.607 \pm 0.017                 
\eqno(8e)$$
and
$$     r_1 - r_2 = ( 2 D + \gamma )/ 3 = 0.549 \pm 0.015             
\eqno (8f)$$  
Defining the left-hand sides as $x_3$ and $x_2$, respectively, we see 
that the equations require $3x_2$ and $x_3$ to coincide. In the 
absence of other information, the most reasonable procedure is to take 
the weighted mean of $3x_2$ and and $x_3$ as the best estimate for $2D 
+\gamma$. Thus, we obtain
$$          2D + \gamma =   1.612  \pm 0.016                 
\eqno (8g)$$
Combined with 
$$          4D + \gamma = 3 (r_1 - r_4) = 3.051 \pm  0.150
\eqno (9e)$$
which follows from Eqs.(9b) and (9c) when $\delta =0$, we obtain the 
solutions
$$         
           D =  0.720 \pm  0.075
\eqno (11e)$$
$$      \gamma =  0.173 \pm  0.154,
\eqno (11f)$$
which, combined with Eq. (8c), yield
$$         F =  0.380 \pm  0.077
\eqno (11g)$$
in substantial agreement with the least-squares fit shown in Table I
to the data (8a)-(8d) under the assumption that SU(3)-breaking in 
baryonic beta-transitions can be represented by the sole parameter 
$\gamma$.}
\end{itemize}

Thus, according to our analysis, the ``spin crisis'' cannot be 
blamed on the effects of SU(3)-breaking and its resolution must be 
sought elsewhere. A critical test of our proposed method of analysis 
requires better data, and application to transitions other than those 
listed in Eqs.(8). In principle, measurement of $g_A/g_V$ for four 
different baryonic beta-decays uniquely determines the couplings $F$ 
and $D$ which survive in the symmetric limit, {\it and} two parameters 
$\gamma$ and $\delta$ representing SU(3)-breaking effects arising from 
mass-splittings. Measurement of additional beta-transitions would 
over-determine the fit and provide a test of the theoretical 
description proposed here to take account of the SU(3)-breaking 
effects which must arise from mass-splittings. It is probable that 
these may become available in the foreseeable future \cite{antos94}. As 
examples of application to possible future data, we calculate $g_A$ for 
$\Sigma^{\pm}\to \Lambda$,   

$$(-g_A)_{\Sigma^{\pm}\to \Lambda}=D+\delta=0.722\pm 0.084,
\eqno (12a)$$
and $g_A/g_V$ for $\Xi^-\to \Sigma^0$,
$$ (-g_A/g_V)_{\Xi^-\to \Sigma^0}=F+D=1.144\pm 0.012,
\eqno (12b)$$   
using the parameters found in Eqs. (11). The equality of the matrix-elements 
for $\Sigma^{\pm}\to \Lambda$ decays is required by the assumption that 
they arise from mirror-conjugate components of the same isovector weak 
current. Similarly, isospin invariance requires the $r$-value for     
$\Xi^0\to \Sigma^+$ to be equal to the one, Eq.(12b), for
$\Xi^-\to \Sigma^0$.

After this note had been written, we came across the paper by the KTeV
Collaboration \cite{ktev01} reporting the decay 
$\Xi^0\to \Sigma^+e^-\bar\nu_e$, which obtained
$$(-g_A/g_V)_{\Xi^0\to \Sigma^+}=1.32\pm 0.20,
\eqno (12d)$$   
which is consistent with our expectation (12b).
 
Note that the value
$$F/D = 0.542 \pm 0.128,   
\eqno (13)$$
deduced from Eqs. (11a) and (11b) is consistent with the representative
value: $F/D=0.575\pm 0.016$ given in \cite{close93,rat99}, and used in the 
spin-structure analyses \cite{e143-smc}. 
 
\bigskip
\vfill\eject

\leftline{\bf 3. Summary}

Effects of flavor-SU(3) breaking, arising from quark mass-differences, 
on hyperon beta-decays have been phenomenologically analyzed. A scheme 
is presented which requires two additional parameters to describe 
the effects of SU(3) breaking in the allowed approximation. Application 
to the available data indicates that such effects have not yet been 
clearly established, and therefore the values of $F$ and $D$ deduced 
from earlier analyses, ignoring SU(3)-breaking effects, do not require 
any drastic revision. One of these parameters $\delta$ is found to be 
consistent with zero, within one standard deviation, at a level of 
accuracy comparable to the $g_A/g_V$ determinations. Test of the 
parametrization proposed in this paper, and demonstration of 
SU(3)-breaking effects in hyperon beta-decays, requires better data, 
and also for transitions beyond those taken into account in our 
numerical analysis.
\bigskip

\leftline{\bf Acknowledgments}

This work was supported in part by the U. S. Department of Energy and 
the Institute of Nuclear and Particle Physics, University of Virginia.
PK thanks the Saha Institute of Nuclear Physics, Calcutta, and XS thanks 
Fudan University and Zhejiang University for hospitality during part of 
this work.

\begin{table}[ht]
\begin{center}
\caption{Parameters describing baryonic beta-transitions, with and without 
 SU(3) symmetry-breaking}
\bigskip
\begin{tabular}{|c|c|c|c|} \hline
Parameters & SU(3)-breaking (I) &  SU(3)-breaking (II) &SU(3) symmetry \\
& ($\gamma\neq 0$,~$\delta\neq 0$)  & ($\gamma\neq 0$,~$\delta=0$)   
& ($\gamma=\delta=0$)  \\
& &[$\chi^2$-fit to $r_{1-4}$]~~~[$r_1$, $r_3$, $r_4$] &[$\chi^2$-fit to 
$r_{1-4}$]~~~~~[$r_1$, $r_2$ only]\\
\hline 
$F$      & $0.402\pm 0.085$ & $0.374\pm 0.075$~~$0.382\pm 0.078$ & 
$0.463\pm 0.025$~~~~~~$0.444\pm 0.023$   
\\
$D$      & $0.742\pm 0.080$ & $0.719\pm 0.075$~~$0.722\pm 0.080$ & 
$0.804\pm 0.025$~~~~~~$0.823\pm 0.023$    
\\
$\gamma$ & $0.123\pm 0.164$ & $0.173\pm 0.150$~~$0.163\pm 0.154$ &  
0~~~~~~~~~~~~~~~~~~~0~~  
\\
$\delta$ & $-0.020\pm 0.024$&0~~~~~~~~~~~~~~0 &  0~~~~~~~~~~~~~~~~~~~0~~    
\\
\hline
$F/D$    & $0.542\pm 0.128$ & $0.520\pm 0.117$~~$0.529\pm 0.123$ & 
$0.576\pm 0.036$~~~~~~$0.539\pm 0.031$   
\\
\hline
\end{tabular}
\bigskip
\end{center}
\end{table}

\begin{table}[ht]
\begin{center}
\caption{Calculated value of $r_j$ under various assumptions.}
\bigskip
\begin{tabular}{|c|c|c|c|} \hline
Transitions & SU(3)-breaking (I)&SU(3)-breaking (II)&SU(3) symmetry \\
& ($\gamma\neq 0$,~$\delta\neq 0$)  & ($\gamma\neq 0$,~$\delta=0$)   
& ($\gamma=\delta=0$)  \\
& &[$\chi^2$-fit to $r_{1-4}$]~~~~[$r_1$, $r_3$, $r_4$] &[$\chi^2$-fit to 
$r_{1-4}$]~~~~~[$r_1$, $r_2$ only]\\
\hline 
$(-g_A/g_V)_{n\to p}$ & input & $1.2667\pm 0.1837$~~~~~~input~~ & 
$1.2670\pm 0.0350$~~~~~~~~input~~~~\\
$(-g_A/g_V)_{\Lambda\to p}$ & " &$0.729\pm 0.146$~~$0.731\pm 0.131$ & 
$0.731\pm 0.026$~~~~~~~~~~~ " ~~~~\\
$(-g_A/g_V)_{\Sigma^-\to n}$ & " & $-0.345\pm 0.106$~~~~~~ input~~ & 
$-0.341\pm 0.035$~$-0.379\pm 0.132$   \\
$(-g_A/g_V)_{\Xi^-\to \Lambda}$& "&$0.25\pm 0.15$~~~~~~~~~~~~~" 
&~~~$0.20\pm 0.03$~~~~~~~$0.17\pm 0.02$ 
\\
\hline
$\chi^2$    & $-$ & ~~0.68~~~~~~~~~~~~~~~0.79 & 1.96~~~~~~~~~~~~~~~8.09    
\\
\hline
$(-g_A/g_V)_{\Xi^0\to \Sigma^+}$& $1.144\pm 0.012$~~
&$1.093\pm 0.106$~~$1.104\pm 0.112$ &$1.267\pm 0.035$
~~~~~~$1.267\pm 0.033$\\
$(-g_A)_{\Sigma^{\pm}\to \Lambda}$& $0.722\pm 0.084$
& $0.719\pm 0.075$~~$0.722\pm 0.080$ &$0.804\pm 0.025$
~~~~~~$0.823\pm 0.023$\\
\hline
\end{tabular}
\bigskip
\end{center}
\end{table}


\begin{references}
\bibitem{note1}
	For the present discussion, we ignore effects associated 
with the $u-d$ mass-difference.
\bibitem{note3}
	Our method of treating SU(3)-breaking is not restricted to this 
approximation.
\bibitem{note4}
	This is not to suggest that Eq. (5) includes $all$ effects of 
SU(3)-breaking.
\bibitem{note6}
	P.~Ratcliffe has independently noted, on the basis of 
  the formulae given in Ref. \cite{skm96}, that only certain linear 
  combinations of perturbative parameters appear in these 
  equations.
\bibitem{note7}
This is because the straight line in the $F$-$D$ plane representing the 
constraint imposed by Eq. (8c) is orthogonal to that corresponding to 
Eq.(8a), whereas the straight line representing Eq.(8b) has only a small 
inclination to the one for Eq. (8a). If one considers bands with widths
corresponding to the errors cited in Eqs. (10a)-(10c), their intersection
extends over a larger range of $F$ in the second case.
\end{references}
\end{document}